\documentclass[aps,pra,twocolumn,superscriptaddress,
              amsmath,amssymb,10pt]{revtex4}

\usepackage{graphicx}
\usepackage[usenames]{color}
\usepackage{epsfig}
\usepackage{verbatim}
\usepackage{listings}
\usepackage{rotating}
\usepackage{amsmath}  
\usepackage{amssymb}
\usepackage{amsfonts}
\usepackage{dcolumn}  

\usepackage{fancyhdr}
\usepackage{bm} 
\usepackage[normalem]{ulem} 

\usepackage{times}

\newcolumntype{d}[1]{D{.}{.}{#1}}  

\newcommand{\mup}{\ensuremath{\mu \mathrm{p} }}
\newcommand{\mud}{\ensuremath{\mu \mathrm{d} }}

\newcommand{\rp}{\ensuremath{r_{\mathrm p}}}
\newcommand{\rd}{\ensuremath{r_{\mathrm d}}}
\newcommand{\rN}{\ensuremath{r_{\mathrm N}}}  

\newcommand{\rstruct}{\ensuremath{r_{\mathrm{struct.}}}}
\newcommand{\rn}{\ensuremath{r_{\mathrm n}}}  

\newcommand{\Ryd}{\ensuremath{R_{\infty}}}

\newcommand{\mred}{\ensuremath{m_\mathrm{red}}}

\newcommand{\ENS}{\ensuremath{E_{NS}}}
\newcommand{\Corr}{\ensuremath{\Delta(n,\ell,j)}}
\newcommand{\CorrP}{\ensuremath{\Delta(n',\ell',j')}}

\newcommand{\lbar}{\mbox{\sout{\ensuremath{\lambda}}\ensuremath{_C}}}

\def\valerrRdTrivial{\ensuremath{2.1422(30)}}

\def\valerrRdDeut{\ensuremath{2.1415(45)}}
\def\DiscrDeut{\ensuremath{3.5 \sigma}}

\begin{document}

\title{Deuteron charge radius and Rydberg constant from spectroscopy data in atomic deuterium}

\def\MPQ{Max--Planck--Institut f{\"u}r Quantenoptik, 85748 Garching, Germany.} 
\def\MAINZ{Johannes Gutenberg Universit{\"a}t Mainz,
QUANTUM, Institut f{\"u}r Physik \& Exzellenzcluster PRISMA,
Staudingerweg 7, 55099 Mainz, Germany.}
\def\LKB{Laboratoire Kastler Brossel, UPMC-Sorbonne Universit\'es, CNRS,
  ENS-PSL Research University, Coll\`{e}ge de France,
  75005 Paris, France.}
\def\LMU{Ludwig-Maximilians-Universit\"at,
Fakult{\"a}t f{\"u}r Physik,
Schellingstrasse 4/III, 80799 Munich, Germany.}
\def\PSI{Paul Scherrer Institute, 5232 Villigen--PSI, Switzerland.}
\def\ETHZ{Institute for Particle Physics, ETH Zurich, 8093 Zurich,
  Switzerland.}

\date{revised version, Oct. 25, 2016}

\author{Randolf~Pohl}                        
\thanks{electronic address: pohl@uni-mainz.de}
\affiliation{\MPQ}
\affiliation{\MAINZ}

\author{Fran\c{c}ois~Nez}
\affiliation{\LKB}

\author{Thomas Udem}
\affiliation{\MPQ} 

\author{Aldo Antognini}
\affiliation{\ETHZ}
\affiliation{\PSI}

\author{Axel Beyer}
\affiliation{\MPQ}

\author{H{\'e}l{\`e}ne Fleurbaey}
\affiliation{\LKB} 

\author{Alexey Grinin}
\affiliation{\MPQ}

\author{Theodor~W.~H{\"a}nsch}
\affiliation{\MPQ}
\affiliation{\LMU}

\author{Lucile Julien}
\affiliation{\LKB} 

\author{Franz~Kottmann}
\affiliation{\ETHZ}

\author{Julian J.\ Krauth}
\affiliation{\MPQ}

\author{Lothar Maisenbacher}
\affiliation{\MPQ}

\author{Arthur Matveev}
\affiliation{\MPQ}

\author{Fran\c{c}ois~Biraben}
\affiliation{\LKB} 


\bibliographystyle{Science}
\begin{abstract}
We give a pedagogical description of the method to extract the
  charge radii and Rydberg constant from laser spectroscopy in regular 
  hydrogen (H) and deuterium (D) atoms, 
  that is part of the CODATA least-squares adjustment (LSA) of the
  fundamental physical constants.
We give a 
  deuteron charge radius \rd\ from D spectroscopy
  alone of \valerrRdDeut\,fm.
This value is
  independent of the measurements that lead to
  the proton charge radius, and
  five times more accurate
  than the value found in the CODATA Adjustment~10.
The improvement is due to the use of a value for the 
  $1S \rightarrow 2S$ transition in atomic deuterium which can 
  be inferred from published data or found in a PhD thesis.
\end{abstract}

\maketitle

\section{Introduction}

For quite a while now, a $7 \sigma$ discrepancy has existed
  between the proton rms charge radius (\rp) determined using electrons
  and muons.
On the one hand, the value from
  laser spectroscopy of the exotic muonic hydrogen atom (\mup),
\begin{equation}
  \label{eq:Rp_mup}
  \rp(\mup) = 0.8409\,(4)\,\mathrm{fm}
\end{equation}
  has been reported by the 
  CREMA collaboration~\cite{Pohl:2010:Nature_mup1,Antognini:2013:Science_mup2}.
On the other hand, the most recent CODATA-2010 ``world average'' value
\begin{equation}
  \label{eq:Rp_CODATA}
  \rp(\mathrm{\mbox{CODATA-2010}}) = 0.8775\,(51)\,\mathrm{fm}
\end{equation}
  has been determined by a self-consistent
  least-squares adjustment (LSA) of the fundamental physical
  constants~\cite{Mohr:2012:CODATA10}.
The discrepancy of $\sim 7\sigma$ between these two values
  has been coined the
  ``Proton Radius Puzzle''~\cite{Pohl:2013:ARNPS,Carlson:2015:Puzzle}.

The CREMA collaboration has just published a value of the deuteron charge
  radius \rd\ from laser spectroscopy of muonic deuterium
  (\mud)~\cite{Pohl:2016:Science_mud}
\begin{equation}
  \label{eq:Rd_mud}
  \rd(\mud) = 2.1256\,(8)\,\mathrm{fm},
\end{equation}
again more than $7 \sigma$ smaller than the CODATA-2010 value of \rd
\begin{equation}
  \label{eq:Rd_CODATA}
  \rd(\mathrm{\mbox{CODATA-2010}}) = 2.1424\,(21)\,\mathrm{fm}.
\end{equation}
However, comparison of the new \rd(\mud) value with the CODATA-2010 value
  may be considered inadequate or redundant,
  because the CODATA values of \rd\ and \rp\ 
  are highly correlated, with a correlation coefficient c(\rp,\rd) = 0.9989
  (see Ref.~\cite{Mohr:2012:CODATA10}, Eq.\,(92)).
This large correlation is the result of the very precisely measured
  isotope shift of the $1S \rightarrow 2S$ transition in
  atomic hydrogen (H) and deuterium (D)
  \cite{Huber:1998:HydrIsoShift,Parthey:2010:PRL_IsoShift},
  which yields a very accurate value for the {\em difference} of the
  (squared) deuteron and proton charge radii~\cite{Jentschura:2011:IsoShift}
\begin{equation}
\label{eq:HDiso}
\rd^2-\rp^2 = 3.82007(65)\,\mathrm{fm^2}.
\end{equation}
One could thus argue that the CODATA deuteron charge radius is 
  larger than the muonic deuterium value
  only because the correlated, and very accurately determined, proton charge
  radius is larger than the muonic hydrogen value.

Here we use the available data on spectroscopy of atomic deuterium to deduce
  a precise value of \rd\
  which does {\em not} depend on
  \rp\ through Eq.\,(\ref{eq:HDiso}).
In our analysis we use a value of the 
  $1S \rightarrow 2S$ transition in atomic deuterium (see Tab.~\ref{tab:D})
  that has not been used by CODATA.
Its value can
  either be inferred from published data or found in a
  PhD thesis~\cite{Udem:PhD}.
This $1S \rightarrow 2S$ value helps improve the accuracy of the
  deuteron charge radius by a factor of five, compared to the
  CODATA Partial Adjustment~10~\cite{Mohr:2012:CODATA10}.

\subsection{CODATA Partial Adjustments}

The final CODATA-2010 recommended values of the fundamental
  constants are deduced in the so-called ``Adjustment~3''.
As detailed in Sec.~XIII.B.2 on page~1577\,ff.\ of the 
  CODATA-2010 report~\cite{Mohr:2012:CODATA10},
  there are additional adjustments
  that use only a subset of the available input data.
``Adjustments 6-12'' are the ones relevant 
  for \rp, \rd\ and the Rydberg constant \Ryd,
  and the results are summarized in Tab.~XXXVIII
  of Ref.~\cite{Mohr:2012:CODATA10}.

These auxiliary Partial Adjustments serve two purposes:
On the one hand, they verify the internal consistency of the CODATA LSA,
  as results from
  different subsets of the data are in good agreement with each other.
On the other hand, these adjustments provide uncorrelated values of
  \rp\ and \rd.
These can then be compared with their muonic counterparts
  to obtain a clearer picture of the 
  issues surrounding the ``proton radius puzzle''. 

For the proton, the value of \rp\ that is deduced from data obtained by
  precision spectroscopy in atomic hydrogen alone
  (omitting both elastic electron-proton (e-p) scattering results 
  and measurements in deuterium)
  is determined in Adjustment~8, 
  see Tab.~XXXVIII of Ref.~\cite{Mohr:2012:CODATA10}:
\begin{equation}
\label{eq:Rp_H_CODATA}
 \rp\mathrm{(H~spectr.,~CODATA)} = 0.8764(89)\,\mathrm{fm}.
\end{equation}
This value is in excellent agreement with Eq.\,(\ref{eq:Rp_CODATA}),
  and only slightly less accurate, see Fig.~\ref{fig:Rp}.
The ``atomic physics'' part of the proton radius puzzle is the
  $4.0\sigma$ discrepancy between Eq.\,(\ref{eq:Rp_mup}) and 
  Eq.\,(\ref{eq:Rp_H_CODATA}).
It is unaffected by the problems that may exist
  in the analysis of e-p scattering data~\cite{Hill:2010:ModelIndepScatt,Lorenz:2012:Closing_in,Kraus:2014:polyFits,Higinbotham:2016:Rp_ep,Horbatsch:2016:ep_scatt}.

The situation is somewhat less favorable for the deuteron charge radius \rd.
The CODATA-2010 value from the full Adjustment~3
  given in Eq.\,(\ref{eq:Rd_CODATA}) is very precise:
  $\rd\mathrm{(CODATA)} = 2.1424(21)$\,fm.
The value from laser spectroscopy of atomic deuterium from Adjustment~10,
  on the other hand, is less so~\footnote{\label{fn:extra_precision}
   The CODATA-2010 report quotes 2.121(25)\,fm, but we
   list all charge radii 
   with 4 decimal figures to make the different accuracies
   immediately obvious.
   The numbers in Eq.\,(\ref{eq:Rd_D_CODATA}) were provided 
   by Barry N.\ Taylor and David B.\ Newell from CODATA/NIST.}:
\begin{equation}
\label{eq:Rd_D_CODATA}
  \rd\mathrm{(D~spectr.,~CODATA)} = 2.1214(253)\,\mathrm{fm}.  
\end{equation}
This value is not accurate enough for a useful comparison with the new
  result from muonic deuterium, see Fig.~\ref{fig:Rd}.

\subsection{The ``missing'' $\boldsymbol{1S\rightarrow2S}$ measurement in D}
The reason for this significantly worse accuracy
  of \rd\ in Eq.\,(\ref{eq:Rd_D_CODATA})
  is the apparent lack of a precise measurement of the $1S \rightarrow 2S$
  transition
  in atomic deuterium.
Only the isotope shift,
  {\it i.e.} the {\em difference} of the
  $1S \rightarrow 2S$ transitions in H and D,
  is used in the CODATA LSA, see
  Ref.~\cite{Mohr:2012:CODATA10}, Tab.\,XI.

This is perfectly valid for the ``full'' CODATA Adjustment~3
  using all available input data.
However, for Adjustment~10 of spectroscopy data in D, the lack
  of a precise value for the $1S \rightarrow 2S$ transition in D
  results in a much larger uncertainty.

In this note we argue that the $1S \rightarrow 2S$ transition
  frequency in atomic deuterium has been measured 
  very accurately by some of the authors at MPQ.
The published
  isotope shifts~\cite{Huber:1998:HydrIsoShift,Parthey:2010:PRL_IsoShift}
  are in fact the calculated differences of the
  measured $1S \rightarrow 2S$ transitions
  in atomic deuterium and hydrogen.

We can thus proceed to deduce a precise
   value of the deuteron radius from deuterium spectroscopy alone,
   combining the 
   $1S \rightarrow 2S$ transition in D,
   measured by some of the authors at MPQ,
   with the $1S \rightarrow 8S$, $8D$, and $12D$ transitions in D,
   measured by some of the authors at LKB.
The new value is five times more precise as the one in 
   Eq.\,(\ref{eq:Rd_D_CODATA}),
   and can be usefully compared to the
   muonic deuterium value of \rd~\cite{Pohl:2016:Science_mud}.


Next we proceed with a pedagogical introduction to the theory of the 
  energy levels in atomic H and D.
We determine the {\em proton} charge radius from atomic hydrogen data
  alone.
Our value is in excellent agreement with the one from CODATA Adjustment~8.
Afterwards we apply the same formalism to the deuterium data.

\section{Energy levels in hydrogen and deuterium}

\begin{table*}[t!]
  \caption{\label{tab:CorrH}
    Values of \Corr\ in kHz for relevant energy levels in atomic hydrogen.
  \Corr\ includes all relevant corrections to the energy levels from 
  fine structure splittings and QED effects.
    The uncertainties are taken from Ref.~\cite{Mohr:2012:CODATA10}, Tab.XVIII.
    They arise mostly from the estimated uncertainty of uncalculated 
    two-loop corrections~\cite{Biraben:2009:SpectrAtHyd}.
    An uncertainty of ``(0)'' denotes ``negligibly small''.}
  \begin{tabular}{l | d   d d  d d}
    \hline
    \hline
    $n$ & \multicolumn{1}{c}{$S_{1/2}$}
        & \multicolumn{1}{c}{$P_{1/2}$}
        & \multicolumn{1}{c}{$P_{3/2}$}
        & \multicolumn{1}{c}{$D_{3/2}$}
        & \multicolumn{1}{c}{$D_{5/2}$} \\ 
    \hline
    1 & -35\,626\,637.5(2.5) \\
    2 & -12\,636\,167.73(31) & -13\,693\,861.67(3)  &  -2\,724\,820.10(3) \\
    3 &  -4\,552\,757.02(9)  &                      &                     & -1\,622\,832.29(0) &  -539\,495.09(0) \\
    4 &  -2\,091\,350.05(4)  &  -2\,224\,408.70(0)  &     -853\,278.87(0) &    -855\,566.25(0) &  -398\,533.10(0) \\
    8 &     -293\,431.56(1)  &                      &                     &    -138\,996.24(0) &   -81\,867.09(0) \\
   12 &                      &                      &                     &     -44\,349.61(0) &   -27\,422.46(0) \\
    \hline
    \hline
  \end{tabular}
\end{table*}
%
\begin{table*}[t!]
  \caption{\label{tab:CorrD}
    Values of \Corr\ in kHz for relevant energy levels in atomic deuterium.
    The caption of Tab.~\ref{tab:CorrH} applies.}
  \begin{tabular}{l | d   d d  d d}
    \hline
    \hline
    $n$ & \multicolumn{1}{c}{$S_{1/2}$}
        & \multicolumn{1}{c}{$P_{1/2}$}
        & \multicolumn{1}{c}{$P_{3/2}$}
        & \multicolumn{1}{c}{$D_{3/2}$}
        & \multicolumn{1}{c}{$D_{5/2}$} \\ 
    \hline
    1 & -35\,621\,512.1(2.3) \\
    2 & -12\,638\,504.55(29) & -13\,696\,839.80(3)  &  -2\,724\,804.25(3) \\
    3 &  -4\,553\,743.34(9)  &                      &                     & -1\,623\,126.89(0) & -539\,493.99(0) \\
    4 &  -2\,091\,828.14(4)  &  -2\,224\,966.95(0)  &     -853\,462.87(0) &    -855\,752.49(0) & -398\,594.64(0) \\
    8 &     -293\,502.94(1)  &                      &                     &    -139\,031.16(0) &  -81\,886.41(0) \\
   12 &                      &                      &                     &     -44\,361.10(0) &  -27\,429.34(0) \\
    \hline
    \hline
  \end{tabular}
\end{table*}

The energy levels in H and D,
 $E/h$ in frequency units [kHz]
 due to the Planck constant $h$,
 can be parameterized~\cite{Biraben:2009:SpectrAtHyd}
 as a function of principal quantum number $n$,
 orbital quantum number $\ell$,
 and total angular momentum $j$, as
\begin{equation}
\label{eq:Etot}
E(n,\ell,j)/h ~ = ~ - \dfrac{c\Ryd}{n^2} \dfrac{\mred}{m_e} ~ ~ + \dfrac{\ENS}{n^3} \, \delta_{\ell 0} ~ ~ + \Corr.
\end{equation}
The first term on the right hand side
 is the famous Bohr result for the energy levels of an 
 electron orbiting an infinitely heavy nucleus $-\Ryd/n^2$, corrected for the 
 leading order nuclear motion by the reduced mass ratio $\mred/m_e$.
Here, \Ryd\ denotes the Rydberg constant,
 $c$ is the speed of light in vacuum, and 
 the reduced mass of the atom with an electron of mass $m_e$
 and a nucleus of mass $m_N$
 is given by 
\begin{equation}
\label{eq:mred}
\mred = \dfrac{m_e \, m_N}{m_e + m_N} = \dfrac{m_e}{1 + \frac{m_e}{m_N}}.
\end{equation}
The mass ratios $m_e/m_N$ are tabulated in Ref.~\cite{Mohr:2012:CODATA10}.

The second term in Eq.\,(\ref{eq:Etot}) is
 the finite nuclear size correction,
 whose leading order is given
 in kHz
 by~\cite{Mohr:2012:CODATA10,Biraben:2009:SpectrAtHyd}
\begin{equation}
\label{eq:NS}
\ENS^{(0)} ~ = ~ \dfrac{2}{3 h} \left( \dfrac{\mred}{m_e} \right)^3
(Z \alpha)^4 m_e c^2
\left( \dfrac{\rN}{\lbar} \right) ^2 .
\end{equation}
Here,
 $\alpha \approx 1/137.036$ 
 is the fine structure constant, 
 $Z = 1$ is the nuclear charge for H and D,
 $\lbar 
 \approx 386.16$\,fm
 is the reduced
 Compton wavelength of the electron,
 and \rN\ is the rms charge radius of the nucleus,
 {\it i.e.} \rp\ for H and \rd\ for D.

The charge radius contribution \ENS\ is significant only for 
 S-states ($\ell = 0$), as indicated by the Kronecker symbol $\delta_{\ell 0}$
 in Eq.\,(\ref{eq:Etot}).

The $1/n^3$ dependence of \ENS\ in Eq.\,(\ref{eq:Etot})
originates from the overlap of the electron's 
 wave function with the extended nuclear charge distribution.
%
%
%
%
%
For our purposes it is convenient to sum $\ENS^{(0)}$ and all other finite 
  nuclear size effects that are proportional to $1/n^3$.
These higher-order nuclear size corrections
 are  $2\times10^{-4}$ of \ENS\, and thus very small,
 see Ref.~\cite{Mohr:2012:CODATA10} Eqs.\,(75), (77) and (78).
We obtain
\begin{eqnarray}
  \label{eq:NS_H}
  \ENS\mathrm{(H)} & = & 1\,564.60 \times \rp^2 \quad \mathrm{kHz/fm^2}, \\
  \label{eq:NS_D}
  \ENS\mathrm{(D)} & = & 1\,565.72 \times \rd^2 \quad \mathrm{kHz/fm^2},
\end{eqnarray}
both with negligible uncertainty on the level of a few Hz/fm$^2$.
For reference, $E_{\rm NS}$ amounts to approx.\ 1100\;kHz and 7100\;kHz
 for the 1S ground state in H and D, respectively.

The third ingredient of Eq.\,(\ref{eq:Etot}), \Corr, summarizes
  all the remaining corrections.
  The largest part of \Corr\ is due to 
  the use of the Dirac equation instead of the simple Bohr formula.
  Other contributions are the 
  fine- and hyperfine-splittings,
  the relativistic, QED, radiative, recoil and Darwin-Foldy corrections,
  finite size corrections for $P$-states, 
  nuclear polarizability,
  and many higher-order contributions.
  These are listed in Sec.~IV.A.1 of 
  Ref.~\cite{Mohr:2012:CODATA10}.

The \Corr\ can be calculated very accurately using the detailed formulas
  found e.g.\ in 
  Refs.~\cite{Eides:2006:Book,Mohr:2012:CODATA10,Horbatsch:2016:Tab_H}.
We list in Tab.~\ref{tab:CorrH} and Tab.~\ref{tab:CorrD} the values
  of \Corr\
  for relevant states in H and D, respectively.
For reference, the sum of all so-called QED corrections,
  included in $\Delta(1,0,1/2)$ of the $1S$ ground state
  in H and D amount to
$8\,171\,663.8 \pm 2.5$\,kHz and
$8\,176\,795.7 \pm 2.3$\,kHz, respectively.
The dominant uncertainties arise from the two-loop
 corrections~\cite{Biraben:2009:SpectrAtHyd},
 and they are responsible for almost all of the uncertainties of the 
 $\Delta(n,\ell,j)$.
The hyperfine splittings of the $1S$ and $2S$ states have been measured
 very accurately~\cite{Cheng:1980:HFS_H1S,Kolachevsky:2004:HFS_H2S,Karshenboim:2005:PPS}.

All constants except \Ryd\ and the radii \rN\ in 
 Eqs.\,(\ref{eq:Etot})-(\ref{eq:NS_D}) are known with sufficient
 accuracy~\cite{Mohr:2012:CODATA10}
 from measurements other than H or D spectroscopy.
This leaves \Ryd\ and \rN\ to be determined from H or D spectroscopy.
Note that we will later only be concerned with {\em transition frequencies}
 between different energy levels,
 so the Planck constant $h$ on the left hand side of Eq.\,(\ref{eq:Etot})
 drops out.

The Rydberg constant \Ryd\ appears in Eq.\,(\ref{eq:Etot}) 
  explicitly only for the 1st (Bohr) term.
This is to emphasize that the full accuracy of $\sim10^{-12}$ is
  required only for the Bohr term, because only the measurements of
  optical transitions
  between levels with different principal quantum number $n$
  are accurate on the $10^{-12}$ level or better, see Tab.~\ref{tab:H}.
These measurements achieve accuracies in the kHz range or better,
  for transitions frequencies of a several hundred THz.

Technically, also the 2nd (finite size) and 3rd ($\Delta(n,\ell,j)$)
  terms contain the Rydberg constant,
  acting as a ``unit converter'' between
  atomic units, used in the calculation of \ENS\ and $\Delta(n,\ell,j)$,
  and the SI unit of frequency, in which the measurements are done.
The accuracy required in the latter terms is much lower,
  on the order of a few times $10^{-8}$.
This becomes obvious from kHz-accuracy required for 
  the \ENS\ (1100\,kHz and 7100\,kHz for H and D, respectively),
  or for the $\Delta(1S)$ ($-35.6\times 10^6$\,kHz).
Thus, these terms do not require the full $10^{-12}$ accuracy in \Ryd.
Instead, one can {\em calculate} \Ryd\ with an accuracy of
  a few parts in $10^{8}$ from the definition
\begin{equation}
\label{eq:Ryd}
\Ryd = \dfrac{\alpha^2 m_e c}{2 h},
\end{equation}
and the values of $\alpha$, $m_e$ and $h$ from measurements other than
  spectroscopy of H or D~\cite{Hanneke:2008,Aoyama:2012:10th_order_g_2,Bouchendira:2011:h_MRb,Stock:2012:WattBalance,Sturm:2014:m_e}.

The CODATA-2010 report lists 24 transition frequencies in H and D that enter
 the LSA, see Ref.~\cite{Mohr:2012:CODATA10}, Tab.\,XI.
We reproduce the most relevant numbers, and a few more,
 in Tabs.~\ref{tab:H}, \ref{tab:iso} and \ref{tab:D}.
In particular, we list several measurements of the $1S\rightarrow2S$ transition
  frequency in D.

Next we introduce the {\em modified} transition frequencies
\begin{equation}
\label{eq:nuCorr}
\widetilde{\nu}[ (n,\ell,j) \rightarrow (n',\ell',j') ]
    = \nu_\mathrm{meas} + \Corr - \CorrP
\end{equation}
where all fine-, hyperfine-, and QED contributions 
 (except for the finite size effect of $S$ states)
 have been removed.
These {\em modified} transition frequencies can then be used to extract 
 \rN\ and \Ryd\ using
\begin{multline}
\label{eq:nuC}
\widetilde{\nu}[ (n,\ell,j) ~ \rightarrow ~ (n',\ell',j') ] ~ = \\
    c\Ryd \dfrac{\mred}{m_e} \left(\dfrac{1}{n^2} - \dfrac{1}{n'^2}\right)
  - \ENS \left(\dfrac{\delta_{\ell 0}}{n^3} - \dfrac{\delta_{\ell' 0}}{n'^3} \right),
\end{multline}
which of course follows from Eq.\,(\ref{eq:Etot}).

\section{Proton radius from hydrogen spectroscopy}
\begin{table*}[t]
  \caption{\label{tab:H}
    Some recent measurements in atomic hydrogen. 
    An asterisk following the reference denotes items considered in 
    the most recent CODATA-2010 report.
    Following our nomenclature, the $2S\rightarrow2P_{1/2}$ transition must be assigned 
    a negative frequency,
    because the final state $(n',\ell',j') = 2P_{1/2}$  is {\em lower}
    than the initial $(n,\ell,j) = 2S_{1/2}$ state.}
  \begin{tabular}{l | l l l l l}
    \hline
    \hline
    \# & $(n,\ell,j)-(n',\ell',j')$ & \multicolumn{1}{c}{$\nu_\mathrm{meas}$ (kHz)} & rel. unc. & Source & Ref. \\
    \hline
    H1 & $2S_{1/2} \rightarrow ~2P_{1/2}$ & \qquad~~~\,
                               -1 057 862(20) & $1.9\times10^{-5}$ &
                  Sussex 1979 & \cite{Newton:1979:H2S2P} *\\
    H2 &                      & \qquad~~~\,
                               -1 057 845.0(9.0) & $8.5\times10^{-6}$ &
                 Harvard 1986 & \cite{Lundeen:1986:LS} *\\
    H3 & $2S_{1/2} \rightarrow ~2P_{3/2}$ & \qquad\quad~
                                9 911 200(12) & $1.2\times10^{-6}$ &
                 Harvard 1994 & \cite{Hagley:1994:FShyd} *\\
    \hline
    H4 & $2S_{1/2} \rightarrow ~8S_{1/2}$ & ~~ 770 649 350 012.0(8.6) & $1.1\times10^{-11}$ &
                     LKB 1997 & \cite{Beauvoir:1997:H2S8SD} * \\
    H5 & $2S_{1/2} \rightarrow ~8D_{3/2}$ & ~~ 770 649 504 450.0(8.3) & $1.1\times10^{-11}$ &
                     LKB 1997 & \cite{Beauvoir:1997:H2S8SD} * \\
    H6 & $2S_{1/2} \rightarrow ~8D_{5/2}$ & ~~ 770 649 561 584.2(6.4) & $8.3\times10^{-12}$ &
                     LKB 1997 & \cite{Beauvoir:1997:H2S8SD} * \\
    H7 & $2S_{1/2} \rightarrow 12D_{3/2}$ & ~~ 799 191 710 472.7(9.4) & $1.1\times10^{-11}$ &
                     LKB 1999 & \cite{Schwob:1999:Hydr2S12D} * \\
    H8 & $2S_{1/2} \rightarrow 12D_{5/2}$ & ~~ 799 191 727 403.7(7.0) & $8.7\times10^{-12}$ &
                     LKB 1999 & \cite{Schwob:1999:Hydr2S12D} * \\
    \hline
    H9 & $1S_{1/2} \rightarrow ~2S_{1/2}$ & 2 466 061 413 187.103(46) & $1.9\times10^{-14}$&
                     MPQ 2000 & \cite{Niering:2000:Hy1S2S} \\
    H10&                      & 2 466 061 413 187.080(34) & $1.4\times10^{-14}$&
                     MPQ 2004 & \cite{Fischer:2004:DriftFundConst} * \\
    H11&                      & 2 466 061 413 187.035(10) & $4.2\times10^{-15}$&
                     MPQ 2011 & \cite{Parthey:2011:PRL_H1S2S}\\
    H12&                      & 2 466 061 413 187.018(11) & $4.5\times10^{-15}$&
                     MPQ 2013 & \cite{Matveev:2013:H1S2S} \\
    H13& $1S_{1/2} \rightarrow ~3S_{1/2}$
                              & 2 922 743 278 678(13) &  $4.4\times10^{-12}$ &
                     LKB 2010 & \cite{Arnoult:2010:1S3S} * \\
    H14&                      & 2 922 743 278 659(17) &  $5.8\times10^{-12}$ &
                     MPQ 2016 & \cite{Yost:2016:1S3S} \\
    \hline
    \hline
  \end{tabular}
\end{table*}
Table~\ref{tab:H} lists 14 transition frequencies in atomic hydrogen. These
 can be separated in three blocks.

%
\subsubsection{Radio-frequency measurements within $\boldsymbol{n=2}$}
\label{sec:H2S2P}
The first block in Tab.~\ref{tab:H}, items H1-H3,
 are radio-frequency measurements of $2S \rightarrow 2P$ transition frequencies
 in H.
Modifying the measured frequencies by $\Delta(2S_{1/2}) - \Delta(2P_{j'})$
  from Tab.~\ref{tab:CorrH},
  each of these three measurements can be used individually to
  determine a value of the proton charge radius \rp\ from Eq.\,(\ref{eq:NS_H})
\begin{equation}
  \label{eq:2S2P}
  \widetilde{\nu}(2S_{1/2} \rightarrow 2P_{1/2}) ~ = ~ \dfrac{1}{8} \, \ENS.
\end{equation}

Each of these three measurements H1-H3 thus yields,
 a value of \rp, listed in Tab.~\ref{tab:Rp}.

As explained above, these three \rp\ values 
 are in fact independent of the {\em exact} value of the Rydberg constant:
The relative uncertainties of the radio-frequency measurements are on the 
 order of $10^{-6}$,
 so only the 6 most significant digits of \Ryd\ enter the calculation.
The ``proton radius puzzle'' could ultimately require a change of
 \Ryd\ by $7 \sigma$, or $10^{-11}$, as explained below. But such a change
 would not affect the \rp\ values obtained from items H1-H3.
 
\subsubsection{Optical measurements between levels with different $\boldsymbol{n}$}
The 2nd block in Tab.~\ref{tab:H}, items H4-H8, lists the 
 five most accurate measurements of
 transition frequencies between the metastable 2S state and 
 higher-$n$ ``Rydberg'' states with $n$=8 or 12.
Because these transitions are between levels with different principal
 quantum number $n$, one has to combine each of these measurements with
 a 2nd measurement to obtain a pair of values for \rp\ and \Ryd,
 using Eq.\,(\ref{eq:Etot}).
Ideally, one combines each of the items H4-H8 with a measurement of the
 $1S \rightarrow 2S$ transition from block 3 in Tab.~\ref{tab:H},
 solving pairs of equations like
\begin{eqnarray}
  \label{eq:1s2s}
  \widetilde{\nu}(1S\rightarrow2S) & = & ~\dfrac{3}{4} c\Ryd\ ~ - ~ ~\dfrac{7}{8} \ENS \\
  \label{eq:2s8s}
  \widetilde{\nu}(2S\rightarrow8S) & = & \dfrac{15}{64} c\Ryd\ ~ - ~ \dfrac{63}{512} \ENS .
\end{eqnarray}
Considering the uncertainties of the experimental values in Tab.~\ref{tab:H}
 and of the \Corr\ in Tab.~\ref{tab:CorrH} 
 one sees immediately,
 that the dominant uncertainty is always given by the
 $2S \rightarrow n\ell$ measurements
 with their experimental uncertainty of the order of $\sim 7$\,kHz.
Several measurements of the $1S \rightarrow 2S$ transition exist
 with uncertainties of much less than 1\,kHz.
Hence one can choose any of the items H9-H12 to reach the same conclusion.

\begin{table}[b]
  \caption{\label{tab:Rp}
    Proton charge radii from hydrogen.
    The row labeled ``CODATA Adjustment~8'' is the value
    using all hydrogen data,
    listed in Ref.~\cite{Mohr:2012:CODATA10}, Tab.\,XXXVIII.
    Also given are the radii from combining the $2S \rightarrow n\ell$
    transitions in H
    with either $1S \rightarrow 2S$ or $1S \rightarrow 3S$.
    All values agree very well.
    ``avg'' denotes the average of all values in the rows above, 
    also considering correlations.
  }
  \begin{center}
  \begin{tabular}{c c c c}
    \hline
    \hline
    \#       &   Transition(s)   & \rp\ [fm]\\
    \hline
    H1       &       $2S \rightarrow 2P_{1/2}$    & $0.9270 \pm 0.0553$ \\
    H2       &       $2S \rightarrow 2P_{1/2}$    & $0.8788 \pm 0.0262$ \\
    H3       &       $2S \rightarrow 2P_{3/2}$    & $0.8688 \pm 0.0354$ \\
    \hline
    H10 + H4 & $1S \rightarrow 2S$ + $2S \rightarrow \;8S_{1/2}$  & $0.8666 \pm 0.0211$ \\
    H10 + H5 & $1S \rightarrow 2S$ + $2S \rightarrow \;8D_{3/2}$  & $0.8789 \pm 0.0204$ \\
    H10 + H6 & $1S \rightarrow 2S$ + $2S \rightarrow \;8D_{5/2}$  & $0.8911 \pm 0.0155$ \\
    H10 + H7 & $1S \rightarrow 2S$ + $2S \rightarrow 12D_{3/2}$ & $0.8551 \pm 0.0222$ \\
    H10 + H8 & $1S \rightarrow 2S$ + $2S \rightarrow 12D_{5/2}$ & $0.8641 \pm 0.0164$ \\[0.4ex]
    \hline
    \multicolumn{2}{l}{\raisebox{0mm}[3mm][1.5mm]{}
      $1S \rightarrow 2S$ (H10) \quad + \hfill all H($2S\rightarrow n\ell$)}  & $0.8747 \pm 0.0091$ & avg.\\
    \hline
    \multicolumn{2}{l}{\raisebox{0mm}[3mm][1.5mm]{}
      $1S \rightarrow 3S$ (H13+H14) + all H($2S\rightarrow n\ell$)} & $0.8780 \pm 0.0108$ \\
    \hline
    \multicolumn{2}{l}{\raisebox{0mm}[3mm][1.5mm]{}
      CODATA Adj.~8 } & $0.8764 \pm 0.0089$ & ~ Eq.\,(\ref{eq:Rp_H})\\
    \hline
    \hline
  \end{tabular}
  \end{center}
\end{table}

We choose the 2004 measurement~\cite{Fischer:2004:DriftFundConst} H10
 with an uncertainty of 0.034\,kHz,
 which was also used in CODATA-2010.
The results are summarized in Tab.~\ref{tab:Rp}.

A trivial weighted average of all individual \rp\ values
 in Tab.~\ref{tab:Rp} yields 
 \rp\ from H spectroscopy alone, of $\rp(\mathrm{H}) = 0.8746 \pm 0.0076$\,fm,
 $4.4 \sigma$ larger than the \mup\ value.
This number is in good agreement with a recent
  evaluation~\cite{Horbatsch:2016:Tab_H}, which finds a
  0.035(7)\,fm, or $4.9\sigma$, difference between H and \mup.

However, relevant correlations exist between the various measurements
 of block~2, see Ref.~\cite{Mohr:2012:CODATA10}, Tab.\,XIX.
These correlations increase the uncertainty of the derived 
$\rp(\mathrm{H})= 0.8747(91)$\,fm.

Alternatively, one can, instead of the $1S \rightarrow 2S$ transition (H10)
  combine the $1S \rightarrow 3S$ transitions (H13 and H14) with
  all $2S \rightarrow n\ell$ transitions. This yields (including correlations)
  $\rp(\mathrm{H'}) = 0.8780(108)$\,fm, in very good agreement with the
  value above, and only slightly less accurate.

\begin{figure}[b!]
  \includegraphics[width=1.0\columnwidth]{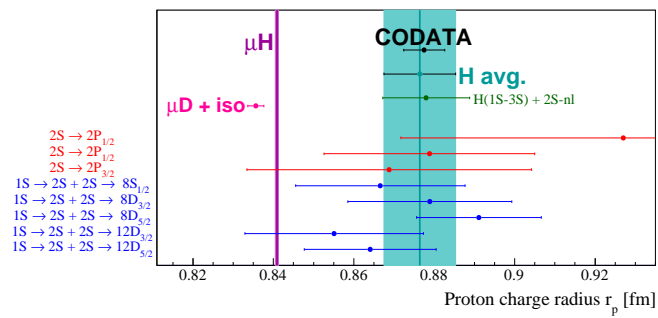}
  \caption{Proton rms charge radii from muonic hydrogen ($\mu$H,
    the stripe includes the uncertainty) and 
    muonic deuterium~\cite{Pohl:2016:Science_mud}
    (``$\mu$D + iso'', obtained using Eq.\,(\ref{eq:HDiso})),
    in comparison with the CODATA-2010 value (Eq.\,(\ref{eq:Rp_CODATA})),
    the value from hydrogen spectroscopy alone (Eq.\,(\ref{eq:Rp_H})),
    and the alternative value from using the $1S\rightarrow 3S$ 
    measurement in hydrogen instead of the $1S\rightarrow 2S$ transition,
    see text. Also shown are the individual values from 
    $2S\rightarrow 2P$ and from combining 
    $1S\rightarrow 2S$ and $2S\rightarrow n\ell$, see Tab.~\ref{tab:Rp}.
    \vspace{2.5ex}\mbox{~}}
  \label{fig:Rp}
\end{figure}

\begin{figure}[b!]
  \includegraphics[width=1.0\columnwidth]{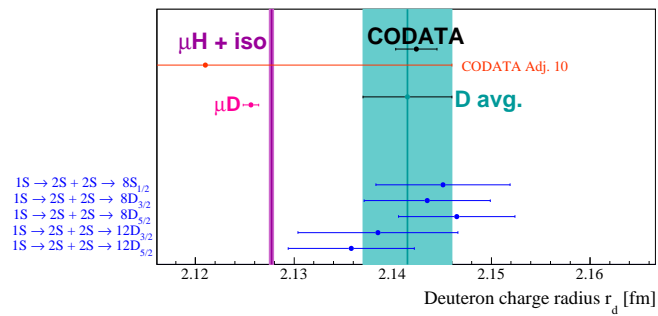}
  \caption{Deuteron rms charge radii from
    spectroscopy of deuterium alone,
    see Tab.~\ref{tab:Rd}, and muonic atoms. Also shown are the CODATA value 
    Eq.\,(\ref{eq:Rd_CODATA}),
    and the value from CODATA Adjustment~10 (Eq.\,(\ref{eq:Rd_D_CODATA}))
    that does not use the
    value for the $1S \rightarrow 2S$ transition in D (see text).
    The value ``$\mu$H + iso''~\cite{Antognini:2013:Science_mup2}
    is obtained from Eq.\,(\ref{eq:HDiso})
    using the proton charge radius from  muonic hydrogen Eq.\,(\ref{eq:Rp_mup}).
    The discrepancy is the same ``proton radius puzzle'' as the one in
    Fig.~\ref{fig:Rp}.
    The new deuteron radius from muonic deuterium~\cite{Pohl:2016:Science_mud}
    ($\mu$D) is \DiscrDeut\ smaller than the
    average value from deuterium spectroscopy (Eq.\,(\ref{eq:Rd_D})).
    }
  \label{fig:Rd}
\end{figure}

A reliable value for the proton rms charge radius deduced from H data alone,
  which takes into account 
  all data in H listed in Tab.\,XI of Ref.~\cite{Mohr:2012:CODATA10},
  as well as the correlations between all input parameters,
  is given in Adjustment~8 of the CODATA-2010 LSA, see
  Ref.~\cite{Mohr:2012:CODATA10}, Tab.\,XXXVIII.
\begin{equation}
  \label{eq:Rp_H}
  \rp\mathrm{(H~spectroscopy)} = 0.8764 (89)\,\mathrm{fm}.
\end{equation}
This value is $4.0\sigma$ larger than the value from muonic hydrogen,
see Fig.~\ref{fig:Rp}.

Considering elastic electron-proton (e-p) scattering data together with
 H spectroscopy,
 as done in Adjustment~9 of the CODATA-2010 LSA, 
 yields \rp(H and e-p) = $0.8796(56)$\,fm,
 which is $6.9\sigma$ larger than the \mup\ value.
This is the ``proton radius puzzle'' between measurements with electrons
 and muonic hydrogen.

\section{Deuteron radius from deuterium spectroscopy alone}
The principle of determining the deuteron radius from deuterium spectroscopy
 is exactly analogous to the one described for hydrogen above.
However, not all measurements were done for deuterium.
Table~\ref{tab:D} lists the relevant deuterium data.

First, we note that there are no radio-frequency measurements of
 $2S \rightarrow 2P$ transitions (i.e.\ no ``block 1'').
Thus there are no ``Rydberg-free'' \rd\ values such as the \rp\ values H1-H3.

Moreover, no measurement of the
 $1S \rightarrow 2S$ transition in ``deuterium only'' 
 is listed in the
 CODATA list of measurements, see Ref.~\cite{Mohr:2012:CODATA10}, Tab.\,XI.
Only the $1S \rightarrow 2S$ isotope shift, {\it i.e.} the difference of the
 $1S \rightarrow 2S$ 
 transition in D and H, is listed there.
We give the two most recent values 
  of the H/D isotope
  in Tab.~\ref{tab:iso}.

This apparent lack of a precise measurement of the
  $1S \rightarrow 2S$ transition in D seems to make it impossible to apply
  the procedure outlined above for hydrogen,
  in which pairs of (\Ryd, \rd) are obtained
  by combining $1S \rightarrow 2S$ and $2S \rightarrow n\ell$ measurements.
CODATA instead performs their Adjustment~10
  of all ``deuterium only'' measurements
  using only $2S \rightarrow n\ell$ measurements
  (plus some much less accurate differences of $2S \rightarrow 4S/D$ and
  $1/4$ of the $1S \rightarrow 2S$ transition~\cite{Weitz:1995:LS_HD},
  which we omit here for brevity).
This has the serious drawback that the ``long lever-arm'' provided by the
  extremely accurate $1S \rightarrow 2S$ transition is lost, which
  is reflected by the large uncertainty of \rd\ obtained in 
  Adjustment~10 of CODATA-2010 of \rd\ = 2.1207(253)\,fm,
  see Eq.\,(\ref{eq:Rd_D_CODATA}).

Several very precise values for the
  $1S \rightarrow 2S$ transition in atomic deuterium exist, however,
  see Tab.~\ref{tab:D}.
The most precise value is obtained by
  simply adding the $1S \rightarrow 2S$ transition frequency in H and
  the $1S \rightarrow 2S$ H/D isotope shift.
Indeed, the published values of the H/D isotope shift are obtained by
  subtracting two frequency measurements of $1S \rightarrow 2S$ transitions 
  in H and D~\cite{Huber:1998:HydrIsoShift,Parthey:2010:PRL_IsoShift}.
For the full CODATA adjustment~3, this choice makes no difference.
However, without the $1S \rightarrow 2S$ transition in D one does not
  obtain the best possible deuteron radius from D spectroscopy in Adjustment~10.

Any frequency measurement is nothing more than a frequency {\em comparison}.
The so-called ``absolute frequency measurements'' are characterized by a
  comparison to a Cs clock~\cite{Bize:2004:FOM}.
Technically, all these comparisons between H and Cs involve intermediate
  comparisons with ``transfer oscillators''.

\begin{table*}[t]
  \caption{\label{tab:iso}
    Some recent measurements of the H-D isotope shift.
    An asterisk following the reference denotes items considered in 
    the most recent CODATA-2010 report.}
  \begin{tabular}{l | l l l l l}
    \hline
    \hline
    \# & Transitions & \quad Frequency (kHz) & rel. unc. & Source & Ref. \\
    \hline
    I1 & D($1S_{1/2} \rightarrow ~2S_{1/2}$) - H($1S_{1/2} \rightarrow ~2S_{1/2}$)
                              & 670 994 334.64(15)  & $2.2\times10^{-10}$&
                     MPQ 1998 & \cite{Huber:1998:HydrIsoShift} \\
    I2 &                      & 670 994 334.606(15) & $2.2\times10^{-11}$&
                     MPQ 2010 & \cite{Parthey:2010:PRL_IsoShift} *\\
    \hline
    \hline
  \end{tabular}
\end{table*}
\begin{table*}
  \caption{\label{tab:D}
    Some recent measurements in atomic deuterium.
    An asterisk following the reference denotes items considered in 
    the most recent CODATA-2010 report.
    Items D9 and D10 are direct measurements
    using a CH$_4$ stabilized He:Ne laser as a transfer oscillator,
    while D11 and D12 have been measured using the $1S\rightarrow 2S$ 
    transition in hydrogen and a hydrogen maser as transfer oscillators.}
  \begin{tabular}{l | l l l l l}
    \hline
    \hline
    \# & $(n,\ell,j)-(n',\ell',j')$ & \multicolumn{1}{c}{$\nu_\mathrm{meas}$ (kHz)} & rel. unc. & Source & Ref. \\

    \hline
    D4 & $2S_{1/2} \rightarrow ~8S_{1/2}$ & ~~ 770 859 041 245.7(6.9) & $8.9\times10^{-12}$ &
                     LKB 1997 & \cite{Beauvoir:1997:H2S8SD} * \\
    D5 & $2S_{1/2} \rightarrow ~8D_{3/2}$ & ~~ 770 859 195 701.8(6.3) & $8.2\times10^{-12}$ &
                     LKB 1997 & \cite{Beauvoir:1997:H2S8SD} * \\
    D6 & $2S_{1/2} \rightarrow ~8D_{5/2}$ & ~~ 770 859 252 849.5(5.9) & $7.7\times10^{-12}$ &
                     LKB 1997 & \cite{Beauvoir:1997:H2S8SD} * \\
    D7 & $2S_{1/2} \rightarrow 12D_{3/2}$ & ~~ 799 409 168 038.0(8.6) & $1.1\times10^{-11}$ &
                     LKB 1999 & \cite{Schwob:1999:Hydr2S12D} * \\
    D8 & $2S_{1/2} \rightarrow 12D_{5/2}$ & ~~ 799 409 184 966.8(6.8) & $8.5\times10^{-12}$ &
                     LKB 1999 & \cite{Schwob:1999:Hydr2S12D} * \\
    \hline
    \hline
    D9 & $1S_{1/2} \rightarrow ~2S_{1/2}$ & 2 466 732 407 521.8(1.5) &  $6.1\times10^{-13}$&
                     MPQ 1997 & \cite{Udem:PhD} \\
    D10&                      & 2 466 732 407 522.88(91) &  $3.7\times10^{-13}$&
                     MPQ 1997 & \cite{Udem:PhD} \\
    D11&                      & 2 466 732 407 521.74(20) & $7.9\times10^{-14}$&
                     MPQ 1998/2000 & H9~+I1 \\
    D12&                      & 2 466 732 407 521.641(25)  & $1.0\times10^{-14}$&
                     MPQ 2010/2011 & H11+I2 \\
    \hline
    \hline
  \end{tabular}
\end{table*}

For example, items I1, D9 and D10 used
  a CH$_4$-stabilized HeNe laser, which was then
  transported to the German Standards Institute PTB
  for comparison with a Cs clock.
In between, a plethora of local oscillators were used in two
  ``frequency chains''~\cite{Udem:PhD}.
More recently, items H9-H12 used a  hydrogen maser
  as a transfer oscillator.
This maser was then compared to a Cs fountain clock~\cite{Bize:2004:FOM}.
 
The isotope shift measurement I2 is a frequency comparison between
  D($1S \rightarrow 2S$) and the same hydrogen maser,
  using GPS calibration.
The maser was then compared to the hydrogen $1S \rightarrow 2S$
  transition.
The practical reason to use hydrogen as an intermediate
  transfer oscillator to the Cs SI clock 
  was that it did not require the
  availability of a primary Cs frequency standard at MPQ.

Thus, we combine items H9 and I1, and H11 and I2, to
  obtain two values for the D($1S \rightarrow 2S$) transition frequency,
  D11 and D12.
This avoids double-counting, 
  because item H10 has been used above to determine the proton radius.
For simplicity, we add the uncertainties linearly, although a
  more rigorous evaluation of the combined uncertainty,
  including all correlations, would certainly yield a smaller uncertainty
  of the D($1S \rightarrow 2S$) transition frequency.

\begin{table}[t]
  \caption{\label{tab:Rd}
    Deuteron charge radii from deuterium.
    The value labeled ``Eq.\,(\ref{eq:Rd_D})'' is our result.
    It is the average of the individual values above it,
    taking into account the known correlations between the $2S \rightarrow n\ell$ measurements.
    The next two values use items D9 and D10, which have not been measured
    using atomic hydrogen as a transfer oscillator (see text).}
  \begin{center}
  \begin{tabular}{c c c c}
    \hline
    \hline
    \#       &   Transitions   & \rd\ [fm]\\
    \hline
    D12 + D4 & $1S \rightarrow 2S$ + $2S \rightarrow \;8S_{1/2}$  & $2.1451 \pm 0.0068$ \\
    D12 + D5 & $1S \rightarrow 2S$ + $2S \rightarrow \;8D_{3/2}$  & $2.1435 \pm 0.0064$ \\
    D12 + D6 & $1S \rightarrow 2S$ + $2S \rightarrow \;8D_{5/2}$  & $2.1465 \pm 0.0059$ \\
    D12 + D7 & $1S \rightarrow 2S$ + $2S \rightarrow 12D_{3/2}$   & $2.1385 \pm 0.0081$ \\
    D12 + D8 & $1S \rightarrow 2S$ + $2S \rightarrow 12D_{5/2}$   & $2.1358 \pm 0.0064$ \\
    \hline
    \multicolumn{2}{c}{\raisebox{0mm}[3.5mm][1.5mm]{}
      D12 + all D($2S\rightarrow n\ell$)} & $2.1415 \pm 0.0045$ & ~ Eq.\,(\ref{eq:Rd_D})\\
    \hline
    \multicolumn{2}{c}{\raisebox{0mm}[3mm][1.5mm]{}
      D9\,+ all D($2S\rightarrow n\ell$)} & $2.1414 \pm 0.0045$ \\[0.2ex]
    \multicolumn{2}{c}{
      D10 + all D($2S\rightarrow n\ell$)} & $2.1411 \pm 0.0045$ \\
    \hline
    \hline
  \end{tabular}
  \end{center}
\end{table}

If one wishes, one could also use the values D9 or D10 which can be found
  in the PhD thesis of Th.~Udem~\cite{Udem:PhD}. 
These values are ``absolute'' frequency measurements
  without the use of hydrogen as a transfer oscillator.

All of the four values D9...D12 are sufficiently accurate to proceed with the
  determination of \rd\ values from combining $1S \rightarrow 2S$ and
  $2S \rightarrow n\ell$ for $n$=8,12, see Tab.~\ref{tab:Rd}.

The trivial weighted average of the values in Tab.~\ref{tab:Rd} is 
  \rd\ = \valerrRdTrivial\,fm,
  {\it i.e.} $5.3\sigma$ larger than the \mud\ value.
Again, however, correlations~\footnote{See
  Ref.~\cite{Mohr:2012:CODATA10}, Tab.\,XIX}
  between the $2S \rightarrow n\ell$ measurements increase
  the uncertainty.
Taking into account these correlations we obtain
\begin{equation}
  \label{eq:Rd_D}
  \rd\mathrm{(D~spectroscopy)} = \valerrRdDeut\,\mathrm{fm}.
\end{equation}
This value is \DiscrDeut\ larger than the new value from muonic deuterium.

For comparison, using, instead of D12, 
  the $1S\rightarrow2S$ measurements
  D9 or D10,  yields 
  $\rd = 2.1414(45)$\,fm and
  $\rd = 2.1411(45)$\,fm, respectively, including the correlations.
The agreement of these three values shows that it is not important which 
  of the available 
  D($1S\rightarrow2S$) measurements is chosen (see Tab.~\ref{tab:Rd}).

Moreover, this ``D spectroscopy'' value is in excellent agreement with the 
  global CODATA value from Adjustment~3,
  \rd\ = $2.1424 \pm 0.0021$\,fm.
This is a strong indication for the internal consistency of CODATA LSA.
This agreement is also evident in the agreement of the Rydberg constants
  from H spectroscopy on the one hand, and D spectroscopy on the other.
This is further discussed in section~\ref{sec:Ryd}.

We emphasize again that this \DiscrDeut\ discrepancy between muonic
  and electronic deuterium spectroscopy measurements is
  as independent as possible of
  any measurement used in the proton charge radius determination.
Correlations may exist because of unidentified systematic shifts in any
  of the electronic or muonic measurements, or missing or wrong theory
  contributions in electronic or muonic atoms.
In the absence of any indication for such an unknown correlation,
  the new \mud\ measurement~\cite{Pohl:2016:Science_mud}
  constitutes an independent discrepancy.

\section{The deuteron structure radius}
  
In the preceding sections we were concerned with hidden or implicit
  correlations between the (CODATA) values of \rp\ and \rd,
  which originate from
  the nature of performing a least-squares adjustment using
  all available input data in H and D.
Here, we could provide values of \rp\ and \rd\ which are
  ``as uncorrelated as possible'' by separating the analysis of H and D
  data.

Physics, on the other hand, is also the source of an explicit
  correlation between \rp\ and \rd, simply because the deuteron
  contains a proton.
The deuteron charge radius is related to the proton charge radius 
  by~\cite{Buchmann:1996:Rdeut,Jentschura:2011:IsoShift}
\begin{equation}
  \label{eq:Dstruct}
  \rd^2 =  \rstruct^2 + \rp^2 + \rn^2 + \dfrac{3 \hbar^2}{4 m_p^2 c^2},
\end{equation}
where $\rstruct = 1.97507(78)$\,fm~\cite{Jentschura:2011:IsoShift}
  is the deuteron structure radius, 
  {\it i.e.} the proton-neutron separation,
  \rn\ is the neutron mean square charge radius
  $<\rn^2> = -0.114(3)$\,fm~\cite{Kopecki:1995:Rneutron,Kopecki:1997:Rneutron},
  and the rightmost term is the Darwin-Foldy correction of 0.0331\,fm$^2$
  due to the
  zitterbewegung of the proton, see~\cite{Jentschura:2011:IsoShift} and
  also the Appendix of Ref.~\cite{Krauth:2016:Annals}.

The 0.8\% smaller deuteron charge radius from muonic deuterium in 
  Eq.~(\ref{eq:Rd_mud}) is very consistent with the 4\% smaller
  proton radius from muonic deuterium Eq.~(\ref{eq:Rp_mup}),
  inserted in Eq.~(\ref{eq:Dstruct}.
This is the reason why the new \rd(\mud) is understood to confirm the
  smaller proton radius from muonic hydrogen~\cite{Pohl:2016:Science_mud}.

%
%

\section{The Rydberg constant}
\label{sec:Ryd}

The correlation coefficient between the proton radius \rp\ and the 
  Rydberg constant \Ryd\ is as large as 0.989 in the CODATA LSA.
Therefore, a change of \rp\ by $x \sigma$ will normally result in a
  change of \Ryd\ by almost the same $x \sigma$.
 
This can be understood by considering Eq.\,(\ref{eq:Etot}), and the accuracy 
  of the measurements in H listed in Tab.~\ref{tab:H}:\\
The accuracy of each of the $2S \rightarrow n\ell$ transitions ($n=8, 12$),
  which determine the accuracy of \Ryd, is about 1 part in $10^{11}$.
As a consequence, the uncertainty of the Rydberg constant in CODATA-2010 is
  about 6 parts in $10^{12}$.
The $1S \rightarrow 2S$ transition, on the other hand, has been measured with an uncertainty
  of 4 parts in $10^{15}$, {\it i.e.} a factor of 1000 more accurately.

A look at Eq.\,(\ref{eq:1s2s}) reveals the correlation: The left side is 
  measured with an accuracy of 0.010\,kHz.
The 1st term on the right side is known only to $\sim 10$\,kHz (3/4 of the
  17\,kHz uncertainty of the CODATA value of $c\Ryd$)~\cite{Mohr:2012:CODATA10}.

\begin{figure}[t!]
  \includegraphics[width=1.0\columnwidth]{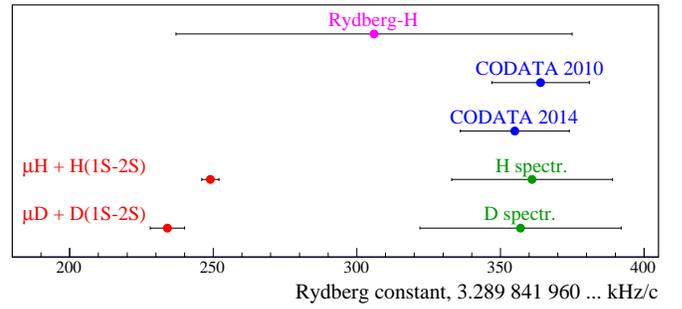}
  \caption{Rydberg constant from
    CODATA-2010 \cite{Mohr:2012:CODATA10}, Eq.~(\ref{eq:Ryd_CODATA10})
    and CODATA-2014 \cite{Mohr:2016:CODATA14}, from
    spectroscopy of regular H and D, Eqs.~(\ref{eq:Ryd_H}) and (\ref{eq:Ryd_D}),
    respectively,
    and from combining the muonic charge radius of the proton and the deuteron
    and the measurement of the $1S \rightarrow 2S$ transition in H and D,
    Eqs.~(\ref{eq:Ryd_mup}) and (\ref{eq:Ryd_mud}), respectively.
    Also shown is the result from spectroscopy of high-lying ($n = 27...30$)
    circular Rydberg states of atomic hydrogen~\cite{deVries:PhD},
    Eq.~(\ref{eq:Ryd_Ryd}).}
  \label{fig:Ryd}
\end{figure}

Adopting the muonic values of \rp\ and \rd\ in \ENS\
  will thus shift the central value of \ENS, 
  which must immediately be compensated by a corresponding
  change in \Ryd\ because of the 1000-fold more precisely determined
  left side of Eq.\,(\ref{eq:1s2s}).
At the same time, the smaller uncertainty of the muonic charge radii
  will yield more accurate values of \Ryd, when combined with
  the electronic $1S \rightarrow 2S$ transitions:
\begin{multline}
\label{eq:Ryd_mup}
\Ryd\ [ H(1S \rightarrow 2S); \rp(\mup) ] = \\
3.\,289\,841\,960\,249\,(3) \times 10^{12}\;\mathrm{kHz/c}
\end{multline}
from electronic and muonic hydrogen~\cite{Antognini:2013:Science_mup2}, and
\begin{multline}
\label{eq:Ryd_mud}
\Ryd\ [ D(1S \rightarrow 2S); \rd(\mud) ] = \\
3.\,289\,841\,960\,234\,(6) \times 10^{12}\;\mathrm{kHz/c}
\end{multline}
from electronic and muonic deuterium~\cite{Pohl:2016:Science_mud}.

The value in Eq.\,(\ref{eq:Ryd_mup}) is in good agreement with the one from
CODATA Adjustment~11, 
\begin{multline}
\Ryd~\mathrm{(Adjustment~11)} = \\
3.\,289\,841\,960\,255\,(4) \times 10^{12}\;\mathrm{kHz/c}
\end{multline}
see Tab.~XXXVIII of Ref.~\cite{Mohr:2012:CODATA10},
which includes \rp\ from muonic hydrogen in the global LSA.
Because of its tiny uncertainty, the  muonic \rp\ value
dominates Adjustment~11, yielding
$\rp~\mathrm{(Adjustment~11)} = 0.84225(65)$\,fm, and this
change of \rp\ is accompanied by a change of \Ryd, as described above.

For reference, the CODATA recommended value of the Rydberg constant is
\begin{multline}
\label{eq:Ryd_CODATA10}
\Ryd~\mathrm{(CODATA-2010)} = \\
3.\,289\,841\,960\,364(17) \times 10^{12}\;\mathrm{kHz/c}
\end{multline}
which is $7 \sigma$ larger.

For completeness,
the values of the Rydberg constant from hydrogen data alone,
taken from CODATA Adjustment~8, is
\begin{multline}
\label{eq:Ryd_H}
\Ryd~\mathrm{(H~spectroscopy)} = \\
3.\,289\,841\,960\,361(28) \times 10^{12}\;\mathrm{kHz/c.}
\end{multline}
The one we deduce from deuterium data alone, including the $1S\rightarrow 2S$
transition is
\begin{multline}
\label{eq:Ryd_D}
\Ryd~\mathrm{(D~spectroscopy)} = \\
3.\,289\,841\,960\,357(35) \times 10^{12}\;\mathrm{kHz/c.}
\end{multline}

A measurement of transition frequencies between high-lying
circular Rydberg states of atomic H, with $n = 27 ... 30$,
which are insensitive to the proton charge radius, yielded~\cite{deVries:PhD}
\begin{multline}
\label{eq:Ryd_Ryd}
\Ryd~\mathrm{(Rydberg-H)} = \\
3.\,289\,841\,960\,306(69) \times 10^{12}\;\mathrm{kHz/c.}
\end{multline}
This result is unfortunately not accurate enough to discriminate the 
muonic and the ``purely electronic'' values, see Fig.~\ref{fig:Ryd}.

New insight into the ``proton radius puzzle'' is expected from several new
atomic physics measurements:
The $2S \rightarrow 4P$ transitions in
  H~\cite{Beyer:2013:AdP_2S4P,Beyer:2013:Conf:ICOLS} will yield an
  independent value of the Rydberg constant.
A new measurement of the classical Lamb shift in H~\cite{Vutha:2012:H2S2P} 
  will yield a proton charge radius that is independent of the exact value 
  \Ryd, see Sec.~\ref{sec:H2S2P}.
Improved measurements of the $1S\rightarrow 3S$ transition in H are
  underway at MPQ and LKB~\cite{Galtier:2015:JPCRD,Fleurbaey:2016:CPEM}.
Measurements of the 1S-2S transition in H-like
  He$^+$ ions~\cite{Herrmann:2009:He1S2S,Kandula:2010:XUV_comb_metrology,Morgenweg:2014:RamseyComb} will,
  when combined with a new value of the alpha particle charge radius
  from muonic helium spectroscopy~\cite{Antognini:2011:Conf:PSAS2010},
  yield a Rydberg constant or test higher-order QED contributions.
The Rydberg constant can also be determined from high-precision spectroscopy of
  molecules and molecular ions of hydrogen isotopes~\cite{Liu:2009:H2Diss,Schiller:2014:MolClock,Dickenson:2013:H2vib,Biesheuvel:2015:HDplus,Karr:2016:HmolIon},
  combined with improved calculations~\cite{Pachucki:2016:H2Schrodinger}.
One-electron ions in circular Rydberg states~\cite{Jentschura:Mohr:fundamental:constants:2008,Tan:NIST:2011} will
  also yield a Rydberg constant free from nuclear radius effects.

As a final remark, we may attribute the small $2.2\sigma$ difference
  between the two Rydberg values using the muonic radii
  (Eq.\,(\ref{eq:Ryd_mup}) and Eq.\,(\ref{eq:Ryd_mud}))
  to the deuteron polarizability contribution~\cite{Pachucki:2011:PRL106_193007,Friar:2013:PRC88_034004,Hernandez:2014:PLB736_344,Carlson:2014:PRA89_022504,Pachucki:2015:PRA91_040503}, summarized in Ref.~\cite{Pohl:2016:Science_mud}.

\section{Conclusions}

The most accurate value of the deuteron rms charge radius
  from laser spectroscopy of regular (electronic) deuterium only
  is \rd = \valerrRdDeut\,fm.
It is obtained using a value for the 
  $1S \rightarrow 2S$ transition in atomic deuterium which can 
  be inferred from published
  data~\cite{Parthey:2010:PRL_IsoShift,Parthey:2011:PRL_H1S2S},
  or found in a PhD thesis~\cite{Udem:PhD}.
Our value is in excellent agreement with the CODATA
  value~\cite{Mohr:2012:CODATA10},
  and only twice less accurate.

In contrast to the CODATA value, the deuteron radius above is
  as uncorrelated as possible to
  measurements that determine
  the proton rms
  charge radius \rp.
The CODATA Adjustment~10, which is also independent of \rp,
   is five times less accurate than the
  value above, because of a more conservative treatment of the deuterium
  $1S \rightarrow 2S$ measurements.

\paragraph*{Note added:} After the submission of this manuscript,
the updated CODATA-2014 paper was published~\cite{Mohr:2016:CODATA14}.
The numbering of the partial Adjustments remained the same.
What was Tab.~XXXVIII in CODATA-10 is now Tab.~XXIX on page~54 of CODATA-14.

The partial Adjustments~8 (H spectroscopy) and 10 (D spectroscopy) yield
  identical values compared to CODATA-10, our Eqs.~(\ref{eq:Rp_H_CODATA}) and 
  (\ref{eq:Rd_D_CODATA}), respectively.
The only new input data is our item H12,
  the 2013 measurement of the $1S \rightarrow 2S$ transition from MPQ.

The change of the recommended values of \rp, \rd, and \Ryd\ (from the full
  Adjustment~3) is exclusively from
  a reassessment of the uncertainty of the electron scattering 
  data~\cite{ArringtonSick:2015:JPCRD}.
None of the conclusions of the present manuscript are changed.

\section{Acknowledgments}
We thank Ingo Sick for insightful comments,
Peter J.\ Mohr, Barry N.\ Taylor and David B.\ Newell from NIST
for providing us with more accurate results of the CODATA LSA which were
very valuable to cross-check our code,
and Kjeld~S.E.\ Eikema for useful remarks.
R.P.\ acknowledges support from 
the European Research Council trough ERC StG.\ 279765,
A.A.\ from the SNF, Projects 200020\_159755 and 
200021\_165854,
and T.W.H.\ from the Max Planck Society and the Max Planck Foundation.
H.F thanks the LABEX Cluster of Excellence FIRST-TF (ANR-10-LABX-48-01),
 within the Program "Investissements d'Avenir" operated by the
 French National Research Agency (ANR) for financial support.


\end{document}